\documentclass{article}
\usepackage{epsfig}

\newfont{\Bbb}{msbm10 scaled 1200} 
\newcommand{\mathbb}[1]{\mbox{\Bbb #1}}
\def\IC{{\mathbb C}}
\def\IR{{\mathbb R}}

\def\Poincare{{Poincar\'e }}

\def\TL{\hfil$\displaystyle{##}$}
\def\TR{$\displaystyle{{}##}$\hfil}

\def\lbldef#1#2{\expandafter\gdef\csname #1\endcsname {#2}}
\def\eqn#1#2{\lbldef{#1}{(\ref{#1})}%
\begin{equation} #2 \label{#1} \end{equation}}
\def\eqalign#1{\vcenter{\openup1\jot
 \halign{\strut\span\TL & \span\TR\cr #1 \cr
 }}}

\def\href#1#2{#2}

\def\sp{ \sin \phi}

\def\u1{{U(1)}}

\def\frac#1#2{{#1\over#2}}

\def\half{\frac12}

\def\d{\partial}

\def\inbar{\,\vrule height1.5ex width.4pt depth0pt}
\def\IC{\relax\hbox{$\inbar\kern-.3em{\rm C}$}}
\def\IR{\relax{\rm I\kern-.18em R}}
\def\IP{\relax{\rm I\kern-.18em P}}

\catcode`\@=11
\def\slash#1{\mathord{\mathpalette\c@ncel{#1}}}
\def\underrel#1\over#2{\mathrel{\mathop{\kern\z@#1}\limits_{#2}}}

\catcode`\@=12

\def\({\left(}
\def\){\right)}
\def\[{\left[}
\def\]{\right]}

\def \sinh{{\rm sinh}}
\def \cosh{{\rm cosh}}
\def \tanh{{\rm tanh}}

\def\sh{{\rm sinh}}
\def\ch{{\rm cosh}}
\def \th{{\rm tanh}}

\usepackage{fortschritte}

\begin {document}

\hfill hep-th/0305137 \vskip .1in\hfill RI-01-03

\def\titleline{Strings in Singular Time-Dependent Backgrounds~\footnote{
Based on talks given by Amit Giveon and Eliezer Rabinovici at the
35'th International Symposium Ahrenshoop on ``The Theory of
Elementary Particles,'' at the Akademie
Berlin-Schm$\ddot{o}$ckwitz, 26-30 August 2002; to appear in the
proceedings.}}
\def\email_speaker{{\tt giveon@vms.huji.ac.il\1ad, eliezer@vms.huji.ac.il\2ad,
asever@phys.huji.ac.il\3ad }}
\def\authors{Amit Giveon\1ad, Eliezer Rabinovici\2ad, Amit Sever\3ad \sp}

\def\addresses{
Racah Institute of Physics, The Hebrew University \\
Jerusalem, 91904, Israel.}

\def\abstracttext{We review the construction of time-dependent
backgrounds with space-like singularities. We mainly consider
exact CFT backgrounds. The algebraic and geometric aspects of
these backgrounds are discussed. Physical issues, results and
difficulties associated with such systems are reviewed. Finally,
we present some new results: a two dimensional cosmology in the
presence of an Abelian gauge field described within a family of
${SL(2)\times U(1)\over U(1)\times Z}$ quotient CFTs.}
 \large
\makefront

\section{Introduction}

The properties of supersymmetry on the worldsheet and in
space-time play an important role in circumventing pitfalls in
string theory. Our universe is not explicitly supersymmetric. Our
universe is time-dependent. Employing the arsenal of familiar
methods to treat strings in time-independent backgrounds in the
cases of time-dependent ones is not straightforward. These are
several of the important problems that string theorists face with
some difficulty.

In this talk we discuss some issues concerning
the motion of strings in time-dependent backgrounds. String theory
can be used reliably to calculate scattering processes involving
gravity only for low energies, $E$, in particular $E\ll 1/g_s$,
where $g_s$ is the string coupling.
If $E\ll 1/l_s$ as well, where $l_s$ is the string length scale, string
theory is validated by reproducing the known results from
General Relativity (GR). When one studies problems which are unresolved
in GR, such as the propagation through
space-like singularities, one would actually wish that the string
theory analysis will deviate, in a subtle but significant manner,
from that of GR. In this way string theory will fulfil its duty
and resolve outstanding problems in GR while adhering to the
correspondence principle.

Is such a behavior possible at all? This actually is the case for
the propagation of strings in the presence of some time-like
singularities. Perturbative and non-perturbative effects do modify
in a subtle way the behavior obtained in a GR framework. A story
in the Talmud helps exemplify the manner in which extended objects
modify point particle problems. During a seminar the following
question came up: Assume one finds a pigeon somewhere, to whom
does it belong? The rule is, finders-keepers, as long as the
pigeon is found further away than some determined distance from
the entrance to an owned pigeon-hole. This seems rather well
defined for point pigeons. A student raised his hand and asked:
``what if the pigeon exhibits its extended object nature? that is
what if one of its legs is nearer than the prescribed distance to
the pigeon-hole but the other is further away?'' The student was
actually ejected from the Yeshiva for asking this question.

In this talk we give a short review on some
of the recent work on strings propagating in time-dependent
backgrounds. The talk has the following structure.
 It starts by reviewing various problems rising in a GR study
of compact cosmology. This follows by suggesting a stringy point
of view on several of these problems. Exact stringy time-dependent
backgrounds are discussed. Special attention is given to the
algebraic, geometrical and dynamical aspects of coset
constructions. The results obtained allow a new point of view on
the problems and are reviewed. The difficulties encountered and
some of the open problems are discussed. These include a study of
singularities; some of the studies in these proceedings are new
(presented in section 7).

\section{Some Questions regarding Cosmologies in General
Relativity}

We start by discussing some of the generic problems encountered by
studying a universe, compact in space, using GR methods.

\begin{itemize}

\item Constructing Time-Dependent Solutions:

It is not immediate to construct time-dependent solutions of the
equations of GR that contain space-like singularities. We will
describe how such solutions can be generated. We will mainly
focus on backgrounds which are in addition exact string
backgrounds. These will be coset models and orbifolds.

\item Observables:

One needs to describe what are the appropriate observables to be
measured in the system. Moreover the universe could even be so
small as to disallow the installation of a classical measuring
device necessary for a quantum measurement.

\item Cauchy data/Singularities:

In GR one studies the evolution of the system as a function of
Cauchy data. In the presence of singularities it may be unreliable
to impose the data and/or the boundary conditions at the
singularity itself. One would rather impose boundary conditions in
regions of space in which the couplings and curvature are small.
In such regions the semi-classical picture may be a useful guide.
Where are such regions? Given appropriate boundary conditions, it is
instructive to follow the scattering of a probe from the
space-like singularity and to calculate the amount reflected
and transmitted. One
would like to know if the question has a precise meaning. Perhaps
 all the information can be encoded in the region up to the
singularity. For black holes there are claims that it is enough to
consider the space-time up to the horizon of the black hole
\cite{Israel}.

\item Entropy \cite{Bousso}:

In the absence of a global time-like Killing vector it is not at
all clear how to define states in GR and how to count them. In
de-Sitter space, arguments have been presented that the total
``number of states'' is actually finite \cite{Banks}. This is done
by analogy to the classical entropy in a black hole. Is there a
stringy microscopic estimate of that entropy?

\item (In)Stability of the Cosmology \cite{hh,hk}:

In GR it is known that there are circumstances under which the
presence of a speck of dust can totally disrupt the global
geometry. Does string theory have anything to add to this?

\end{itemize}

\section{Stringy Attempts to Address the Questions}

\begin{itemize}

\item Constructing Time-Dependent Solutions:

Time-dependent backgrounds containing space-like singularities
have been constructed and studied using several methods.

\begin{enumerate}

\item Exact String Solutions \cite{egkr,egr,ess,nek,cc,gp,bs,nw,lms1,lms2,ckr}:

Exact conformal field theories which contain space-like
singularities have been constructed. The advantage of such
constructions is that they are exact as far as $\alpha'$
corrections are concerned. This allows for example the exact
calculation of vertex operators using algebraic methods without
the need to resort to a classical geometrical picture. This was
done for a class of coset models $ G/H \times M$ where $M$ is an
attendant manifold to a time-dependent four dimensional cosmology.
This was also done for a class of orbifolds of Minkowski space.
The results are mainly for the tree approximation as far as string
perturbation theory is concerned. We review in some detail the
results of such constructions.

\item S-Branes:

These are proposed solutions which are supposed to be in some
sense Dirichlet branes with a fixed coordinate in the time
direction. In a manner the idea that such objects may indeed be
exact solutions are inspired by the presence of the Euclidean
instantons. For a review and references see J.~Walcker's talk.

\item Holography Inspired Relations:

A non-perturbative description of the properties of strings in
time-dependent backgrounds is desirable. Such a description was
made possible in those systems in which the property of Holography
was essentially shown to exist. If one assumes that topology is
valid also in some time-dependent background settings, one may
wish to claim that this holographic description is obtained by
simply making the world-volume of the holographic dual theory be
the appropriate boundary time-dependent background. The
non-perturbative description of the string theory being
essentially a boundary field theory on an appropriate
time-dependent background. The analysis of this field theory is
yet to be done. Aspects of holography and its applications are
discussed in V.~Balasubramanian's talk, (see references there).

\item Solutions Obtained by Double Wick Rotating GR Solutions
which Contain Time-Like Singularities \cite{afhs}:

These are solutions to
Einstein's equations but are not exact string backgrounds. They
are useful to study phenomena such as particle creation which
occurs in time-dependent backgrounds as well as to capture
important features of the stability of the system.

\item RG Flow Induced by Unstable Configurations \cite{sen}:

Tachyonic decay in the open string sector of D-branes and in some
closed string settings can also be formulated as a time-dependent
background. It has been suggested that this has implications in
realistic cosmologies. It is a useful setup for studying time
independent backgrounds. These important studies are not reviewd here.

\end{enumerate}

\item Observables:

One needs to describe what are the appropriate observables to be
measured in the system. In the string theory these may be the
BRST-invariant operators. Moreover in string theory the S matrix
is the observable. In a compact universe it is not clear where to
place the asymptotic scattering states. This is resolved in a
class of models in a rather surprising manner. Some exact string
backgrounds turn out to contain in addition to compact
time-dependent cosmologies also static regions which extend all
the way to spacial infinity and are called ``whiskers''
(see figure 1).~\footnote{In section 7 we shall discuss instead
examples where the static regions are compact and singular.}

\begin{figure}[h] \begin{center} \epsfxsize = 4.5in \epsffile{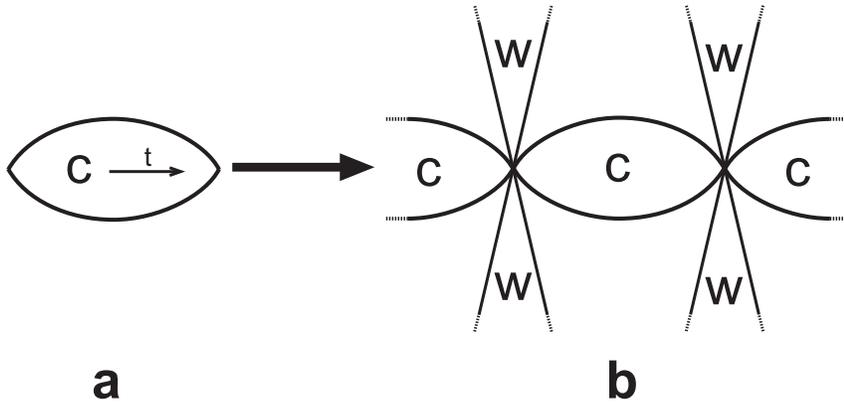}
\caption{a: A compact cosmology (C) in GR. b: The same compact
cosmology repeats itself in string theory and comes with
additional regions -- whiskers (W).} \label{fig 1} \end{center}
\end{figure}

Moreover, the system is weakly coupled at the boundary of these
regions. The accelerators and detectors can thus be placed there to produce
and detect scattering states. There is a possible fly in the ointment; these
regions contain closed time-like curves and may contain in some
cases time-like singularities as well.

\item Cauchy data/Singularities:

Once a time-independent weakly coupled region exists it is natural
to use it to enforce boundary conditions and to give the Cauchy
data. The behavior of the system at the singularity is determined
by the behavior at these better understood regions. In fact, the
reflection coefficient of a given partial wave vertex operator can
be calculated and is found to be smaller than one \cite{egkr}. The
scattering can be shown to be unitary. For the orbifold
backgrounds also the scattering of two vertex operators has been
calculated \cite{lms1,lms2}. The results indicated problems; these
are discussed later.

\item Entropy:

In string theory the entropy can be estimated by counting the
BRST-invariant states. In a class of models the number is
significantly depleted relative to other non-cosmological systems
\cite{egkr}.

\item (In)Stability of the Cosmology \cite{law}:

There are at least two types of stability problems in such
systems: short and long time scale problems. Consider first the
short time scale problem. In the presence of a
space-like singularity, classically the energy density increases
without bounds as the projectile approaches the singularity which
is concentrated at zero volume; this would be suggested even if
the energy is conserved. If it is blue shifted as well as one
approaches the singularities the situation would only be worsened.
In the orbifold models this was pointed out from several points of
view which had the common feature of closed time-like curves and
space-like singularities. A prototype of such a universe is the
Misner Universe. Its metric is given by
 \eqn{metricmisner}{ds^2=-dt^2+t^2dx_1^2+dx_2^2+dx_3^2}
with the identification:
 \eqn{compactx}{(t,x_1,x_2,x_3)=(t,x_1+na,x_2,x_3)}
After a change of coordinates:
 \eqn{ycor}{\eqalign{y_0 &\equiv t\sh(x_1) \quad y_2 \equiv
 x_2 \cr y_1 &\equiv t\ch(x_1) \quad y_3\equiv x_3}}
the metric becomes:
 \eqn{newmetric}{ds^2=-dy_0^2+dy_1^2+dy_2^2+dy_3^2}
with the identification (see figure 2):
 \eqn{compacty}{(y_0,y_1,y_2,y_3)=(y_0\cosh(na)+y_1\sinh(na),y_0\sinh(na)
+y_1\cosh(na),y_2,y_3)}
 \begin{figure}[h] \begin{center} \epsfxsize = 3.5in \epsffile{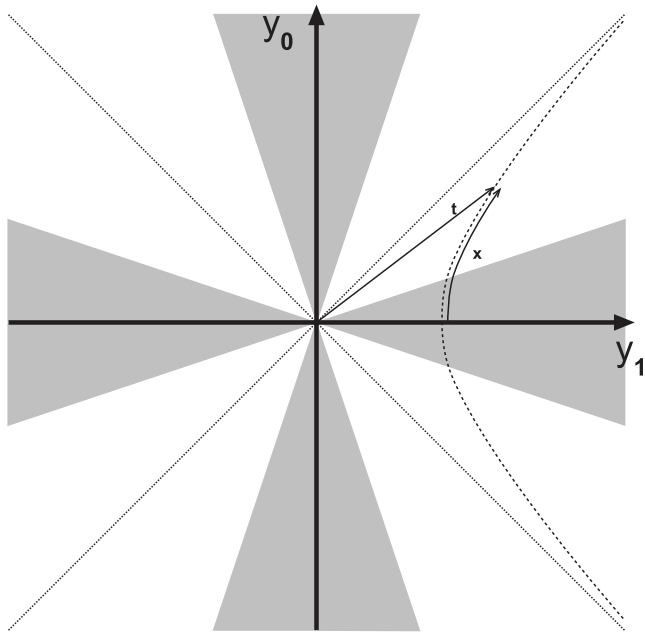}
\caption{The extended Misner universe; a fundamental domain is
marked.}\label{fig 2}
 \end{center} \end{figure}
 The identification has a fixed submanifold given by:
 \eqn{fixedpoints}{(y_0,y_1,y_2,y_3)=(0,0,y_2,y_3)}
The resulting orbifold contains close time-like curves.
 A free field type calculation of the expectation value of the
energy-momentum tensor as a function of the distance to the above
singularity, $t_s$, is singular and given by \cite{hk}:
 \eqn{Tmunu}{\langle T_{\mu\nu}\rangle =-i \lim_{y_1\rightarrow
y_0}\({2\over 3} \nabla_\mu\nabla_\nu ...\)G^{\mu\nu}(y_0,y_1)
={a\over t_s^4} \rm{diag}(1,-3,1,1)}
where $a$ is a system dependent constant. We will discuss more
aspects of this singularity. As mentioned, the system is also
threatened by a long time scale instability. This may occur for
systems with a discrete spectrum and a finite number of ``states''
or at least a finite thermal entropy. Compact cosmologies may be
systems of this nature. For the AdS case the instability starts
when the contribution of the eternal black hole masterfield,
embedded in AdS, is superceded by that of non-dominant thermal AdS
masterfield \cite{maldacena}. Recall that these two masterfields
have different topologies \cite{whp}.
These instabilities were indicated from GR arguments.
On the other hand, in time-independent string
backgrounds we are familiar with two types of instabilities:
perturbative ones -- tachyons -- which face no barrier in their
decay, and non-perturbative instabilities (such as the formation
of small black holes from flat Minkowski space at finite
temperature) which do need to overcome a barrier.
\end{itemize}

\section{Exact Time-Dependent Backgrounds}
Two types of exact time-dependent backgrounds containing
space-like singularities have been considered. Both have bosonic
and fermionic versions with the respective central charge $c=26$
and $c=15$. The first type are $G/H$ cosets. In particular the
coset \cite{egkr,nw}:
 \eqn{spacee}{{SL(2,\IR)_k\times
SU(2)_{k}\over U(1)\times U(1)}\times M}
In this case an identification is done on the group elements of
$G$ by an operation with group elements of $H$ in the following
manner:
 \eqn{gauge}{{G/H}: \quad(g,g')\rightarrow
(e^{\rho\sigma_3}ge^{\tau\sigma_3},e^{i\tau\sigma_3}g'e^{i\rho\sigma_3})~,
\qquad (g,g')\in SL(2)\times SU(2)} The second type are orbifolds.
A particular identification is given by \cite{lms1}:
 \eqn{orbifold}{{G/\Gamma}: \quad (x^+,x,x^-)\rightarrow (x^+,x+2\pi
x^+,x^-+2\pi x + \half(2\pi)^2x^+ )} $G$ is actually $1+2$
Minkowski space in this case. Both identifications have fixed
points but they are of a different nature. The form of the
manifold \orbifold\ is ${R^{1,2}\over \Gamma}\times M$, where
$\Gamma$ is a finite parabolic subgroup of the \Poincare
symmetries of Minkowski space. The metric of the
${R^{1,2}\over \Gamma}$ factor is:
 \eqn{metric}{\ ds^2=-2dx^+dx^-+dx^2}
with the identification \orbifold. The fixed plane is $(0,0,x^-)$;
it is the origin of the light-like orbifold singularity. The surface
$x^+=0$ has closed null-like curves.

The metric of \metric, \orbifold\ can be written as:
\eqn{metricc}{ds^2=-2dy^+dy^-+\(y^+\)^2dy^2,\quad y\sim y+2\pi n}
where \eqn{yspace}{y^+ = x^+~, \qquad y = {x\over x^+}~,\qquad y^-
 = x^- - \half{x^2\over x^+}}
This is similar to the Misner space \metricmisner, \compactx.
At $y^+=0$ there is a light-like big bang/crunch singularity, and near
$y^+=0$ the space is non-Hausdorff.

Different orbifolds discussed in the literature
\cite{cc,nek,lms1,law,orbi,fs}\ are, for instance: $t\rightarrow
-t$, $t\rightarrow t+2\pi$, and other parabolic and hyperbolic
orbifolds. In the above cases $G$ was taken to be the \Poincare
symmetry group of the $d+1$ Minkowski space, and thus is a known
CFT. Different types of orbifolds are the BTZ black holes in which
case $G$ is the group manifold $SL(2,\IR)$. The orbifolds
$G/\Gamma$ have physical twisted sectors in most cases.

Some of the systems are supersymmetric. Supersymmetry prevents
among other things particle production, removes tadpoles, tachyons
and makes the one loop ``cosmological constant'' vanish. At this
stage it is not clear if supersymmetry really plays an important
role in allowing a string perturbative analysis of the theory. The
main positive result was that for a class of these supersymmetric
backgrounds the two and three point functions were calculated and
were found to be well behaved despite of the semiclassical
space-like geometric singularity. The worse is yet to come:
divergences appear in higher point, string tree-level correlators,
in some models \cite{lms1}.

We now turn to discuss some general aspects of coset CFTs,
and coset models leading to time-dependent backgrounds in particular.

\section{Algebraic, Dynamic and Geometrical Aspects of Coset
backgrounds}

\begin{enumerate}

\item Algebraic Aspects of Coset Models \cite{bh_go}:

The special feature of the algebraic treatment of coset CFTs is that it
allows one to obtain the exact vertex operators and in principle
any $n$-point function without the need to worry about the
semiclassical geometrical description of the system. Of course,
singularities in the geometrical description will either be cured
or come back to haunt us in the algebraic description. In order to
appreciate the string backgrounds constructed out of cosets
consider first the WZW models. From an algebraic point of view one
is given a group $G$ from which one can construct first an
affine Lie algebra and then representations of the Virasoro
algebra. Algebraically it is simpler to consider worldsheet
chiral algebras. In particular the operator product expansion
(OPE) defining the affine Lie algebra of $G$ is:
 \eqn{OPE1}{J^a(z_1)J^b(z_2)={k\delta ^{ab} \over z^2_{12}}
 +{i{f^{ab}}_c J^c(z_2)\over z_{12}}}
where $J$ are chiral worldsheet currents and $k$ is the level of
the affine algebra. A similar OPE holds for the anti-chiral
currents, $\bar{J}^a$. The chiral component of the energy-momentum
tensor is constructed out of the currents $J$ by:
 \eqn{T}{T\equiv{:J^aJ_a: \over k+h}}
where $h$ is the dual coxeter number of the Lie group. For example,
$h=k+2$ for $SU(2)$ and $h=k-2$ for $SL(2,\IR)$.
The OPE of two $T$'s is:
\eqn{OPE2}{T(z_1)T(z_2)\sim {c\over
 z^4_{12}}+{2T(z_2)\over z^2_{12}}+{\d_2 T\over z_{12}}}
where $c={k\cdot dimG\over k+h}$ is the Virasoro central charge.
For $SU(2):\,\, c={3k\over k+2}$, and for $SL(2,\IR):\,\, c={3k\over k-2}$.
The algebraic coset $G/H$ also induces a representation of the
Virasoro algebra. Given a group $G$ with a subgroup $H$, one constructs
the following energy-momentum tensor:
 \eqn{GmodH}{T_{G/H}\equiv T_G - T_H={:jj:_G\over k_G+h_G}-
{:jj:_H\over k_H+h_H}}
where $c(G/H)=c(G)-c(H)$. For example,
$c(SU(2)/U(1))={3k\over k+2}-1$, $c(SL(2)/U(1))={3k\over k-2}-1$
(when the $U(1)$ is null replace $-1 \rightarrow -2$).
Can a physical meaning be attributed to the minus sign responsible
for the decrease in the central charge? The answer is yes and we
will now turn to discuss it.

\item The Physics of coset models:

We will describe the coset models as an infra-red limit of two
dimensional confining gauge theories. To obtain that consider
first a conformal field theory: $L=L_{CFT}$ with some Virasoro
central charge $c_{UV}$. Next add to the theory a relevant
operator, or add an operator which is classically marginal but not
truly marginal. This operator should have an asymptotically free
flow. The resulting Lagrangian is: $L=L_{CFT}+\sum_i g_iO_i \quad
(``{\rm relevant}")$. The ultra-violet (UV) limit of the theory is
described by the Lagrangian $L_{CFT}$. The infra-red (IR) limit of
any field theory is a conformal theory. Generically it has a
vanishing central charge, $c_{IR}=0$. In a unitary theory:
$c_{IR}<c_{UV}$ \cite{zamo}. As a first example consider as the UV
theory a free Dirac fermion and modify it by adding the relevant
mass operator. The UV Lagrangian is:
\eqn{lcft1}{L_{CFT}=\bar{\psi}\relax{\rm \partial\kern-.6em
/}\psi} This system has a $c=1$ Virasoro central charge. The full
Lagrangian is not conformal and given by:
\eqn{lcft2}{L=\bar{\psi}\relax{\rm
\partial\kern-.6em /}\psi+m\bar\psi \psi}
 This is an exactly solvable system. It has a mass gap which
becomes infinite in the extreme IR limit. Only the ground state(s)
survive in this limit. The topology of the worldsheet may allow
several vacua which may constitute a topological field theory. In
the absence of non-trivial topology there is a single vacuum
state. Be that as it may, the IR theory has $c=0$ which is smaller
than $c=1$.
In the second example one gauges the full global $U(1)$ that the
free Dirac Fermion Lagrangian posses. This is described by:
 \eqn{lfd}{L=\bar{\psi}\relax{\rm\partial\kern-.6em /}
 \psi+J_\mu A^\mu + {1\over g^2}F^2_{\mu\nu}}
where $ A^\mu $ is the vector potential. The coupling is
dimensionfull, with dimensions of a mass, thus the gauging results
in adding a relevant perturbation to the free Dirac Fermion.
This is the massless Schwinger model. The potential between two
heavy charged sources is confining, $V(r)\sim r$. In fact, with
massless Fermions the theory (super) confines. Only one bound
state is formed and it is a local object having for example no
dipole moment (in the case of massive Fermions they do have a
dipole moment and several bound states may form). In the extreme
IR the gauge coupling becomes infinite and only the ground state
remains (up to perhaps topological issues mentioned before). To
see this explicitly recall that the bosonized version of the
Lagrangian is:
\eqn{bosonl}{L=\half (\d_\mu\sigma)^2+{m^2\over 2}\sigma^2}
where $\sigma$ is a superlocal bound state -- a free massive scalar
with $m\sim g$. In the IR $g,m\rightarrow\infty$ hence all excitations
decouple. It leaves behind only the vacuum state(s), a $c=0$ IR theory.
As in the massive Dirac Fermion case
$c_{UV}=1\rightarrow c_{IR}=0$. Moreover, in the IR
$g\rightarrow\infty$, thus in the IR the term
$F^2_{\mu\nu}\over g^2$ disappears from the Lagrangian \lfd,
leaving just the terms linear in the gauge potential.
The potential is reduced to a
Lagrange multiplier and this is the actual Lagrangian
representation of the coset model:
\eqn{lir}{L_{IR}=L_{CFT}+J_\mu A^\mu}
The third example is the Lagrangian representation of a special
class of coset models. For simplicity we consider here the group
$G$ to be $SO(N)$. In that case the $WZW$ model can be expressed
as a free Fermion system. Its Lagrangian is:
 \eqn{lcft}{L_{CFT}=L_{Free-Fermions}(G)}
Consider gauging the full $SO(N)$ group. This gives:
 \eqn{lff}{L=L_{FF}+J_\mu(G)A^\mu(G)}
This is the Lagrangian of the coset $G/G$ which has a vanishing
Virasoro central charge $c=0$. All the particles obtain a mass as
confined bound states which decouple in the extreme IR. The last
example is the general coset \cite{subgauge}. It is obtained by
gauging a subgroup $H$ of $G$: \eqn{lg}{L=L(G)+J_\mu (H)A^\mu (H)}
This is a schematic description of WZW system of the group $G$
gauged by a subgroup $H$. The exact form appears below. The
resulting central charge decreases as: $c(G/H)=c(G)-c(H)$. The
decrease in central charge is traced to the partial confinement,
that of $H$ singlets.

\item Geometrical Aspects of Coset Models:

The geometrical treatment starts, as the algebraic one did, by first
considering the affine Lie symmetry.
One writes down a Lagrangian which has a conformal as well as affine
symmetry. That Largangian leads to a Hamiltonian which once
diagonalized has eigenfunctions which form representations of both
the conformal group and the affine Lie algebra. One is familiar in
string theory with a geometrical description of the target space
in which the string propagates by a non-linear $\sigma$ model
(NLSM). Given the group $G$ one may consider writing down the
appropriate NLSM by parameterizing the group member $g\in G $ as $
g=e^{ix^\alpha T_\alpha}$ and forming the NLSM
$L=G_{\alpha\beta}\d x^\alpha \bar\d x^\beta$ where $
G_{\alpha\beta}$ is the appropriate metric on the group manifold.
This however is usually not a CFT. In fact, the beta-function of
the couplings $G_{\alpha\beta}(x)$ obey (to leading order in
$\alpha'$):
 \eqn{betafunc}{R_{\mu\nu}\sim\beta_{G_{\mu\nu}}
\begin{array}{ll} >0&
\rm{For\ a\ non-compact\ group\ the\ theory\ is\ IR\ free}
\cr <0 & \rm{For\ a\ compact\ non-abelian\ group\ the\ theory\ is\
UV\ free} \cr =0 & \rm{The\ theory\ is\ already\ conformal\ for\
an\ abelain\ group} \end{array}}
 To get a CFT for any group $G$ the Lagrangian, $L$, needs
to be modified. An antisymmetric field $B_{\alpha\beta}$ is added
to the geometric target space description to make the theory
conformal (this operation is sometimes called to parallize the
manifold). The new Lagrangian is:
\eqn{newL}{L=g^{ij}G_{\alpha\beta}\d_ix^\alpha\d_jx^\beta+\epsilon^{ij}
B_{\alpha\beta} \d_ix^\alpha\d_jx^\beta} The geometrical
description of the coset is given by gauging the subgroup $H$ of
$G$. The $H$ locally gauge invarinat Lagrangian is given by:
 \eqn{SWZW}{\eqalign{S[g]&={k \over {4
\pi}}\int_{\Sigma}dz^2\[Tr(g^{-1}\partial g g^{-1}\bar \partial
g)+ A^a_z A_{\bar{z}}^a +(A_{\bar{z}}^a Tr(g^{-1}\d gT^a)
 +c.c)\] \cr &+{ik \over 12\pi} \int_{B} Tr (g^{-1}dg)^3 }}
This is the full form of the gauged Lagrangian. The NLSM
describing the coset can be obtained by integrating out the gauge
fields, $A_{\mu}^a$. The integral is a guassian one. Actually one
needs to take note of the measure as well. This results in a
modified NLSM, the metric and the antisymmetrical tensor change
and a new background field -- the dilaton $\Phi$ -- emerges:
$(G,B)_{WZW}\rightarrow(G',B',\Phi)_{coset}$ \cite{2dbh}. This
geometric description is exact only for large $k$. Suggestions for
obtaining ``exact'' geometrical backgrounds have been put forward.
In any case the geometric data provides a semi-classical picture.
The algebraic treatment allows precise calculations. In abelian
quotients the two and three point functions are identical to their
corresponding ones in $G$. Those functions are smooth on the group
and therefore also on the coset even if the coset has
singularities. This ends the general discussion we now turn to the
background at hand.

\end{enumerate}

\section{Cosmology and Whiskers \cite{egkr,nw}}

{}From the algebraic point of view the system is the coset
 \eqn{space2}{{SL(2,\IR)_k\times SU(2)_{k}\over
 U(1)\times U(1)}\times M}
As the original motivation is to construct a compact manifold with
space-like singularities one is led to examine the semi-classical
geometry induced by the algebraic coset. This geometry
includes time-dependent cosmological regions and whiskers, as mentioned.
In the cosmological regions the metric, dilaton and B field are
\cite{egkr}:
 \eqn{metrici}{{1\over k}ds^2=-d\theta_1^2+d\theta'^2+{\cot^2\theta'\over
{1+\tan^2\theta_1 \cot^2\theta'}}d\lambda_-^2 + {\tan^2\theta_1
\over { 1+ \tan^2\theta_1 \cot^2 \theta'}}d\lambda_+^2}
 \eqn{bfi}{B_{\lambda_+,\lambda_-}={k \over {1+\tan^2\theta_1 \cot^2\theta'}}}
 \eqn{dili}{\Phi=\Phi_0-{1\over 2}\log(\cos^2\theta_1 \sin^2 \theta' +
 \sin^2\theta_1 \cos^2\theta')}
where $\lambda_{\pm}\in [0,2\pi)$, and $\theta_1$ and $\theta'$
vary in the interval $[0,{\pi\over 2}]$. In the
whiskers:~\footnote{The geometric data obtained is valid in the
large $k$ limit. For the bosonic string there are known $1/k$
corrections \cite{bs,1overk}. The exact background sometimes has a
different singularity structure. For fermionic strings, the
semiclassical background is expected to be a solution to all
orders in $1/k$. The dilaton $\Phi$ is normalized such that the
string coupling is $g_s=e^{\Phi}$.}
 \eqn{metrico}{{1\over k}ds^2=d\theta_2^2+d\theta'^2+{\cot^2\theta'\over
{1-\tanh^2\theta_2 \cot^2\theta'}}d\lambda_+^2 - {\tanh^2\theta_2
\over { 1- \tanh^2\theta_2 \cot^2 \theta'}}d\lambda_-^2}
 \eqn{bfo}{B_{\lambda_+,\lambda_-}={k \over {1-\tanh^2\theta_2 \cot^2\theta'}}}
 \eqn{dilo}{\Phi=\Phi_0-{1\over 4}\log(\cosh^2\theta_2 \sin^2 \theta' -
 \sinh^2\theta_2 \cos^2\theta')^2}
where here $\theta_2\in [0,\infty),\ \theta'\in [0,{\pi \over
2}],\ \lambda_{\pm} \in [0,2\pi)$.

The geomerty is somewhat complex, and follows a list of features
which play a role in addressing the questions posed in the
beginning of the talk:

\begin{itemize}

\item Time-Dependent Comological Regions:

The geometry \metrici\ indeed describes a compact cosmological
region, it is denoted by C in figure 1. In fact, the exact string
background contains an infinite number of such C regions, (for a
partial list of earlier works see \cite{tdcr}). Each C
region can be viewed as starting from a mild big bang in which
only one of its spacial coordinate directions
degenerates, and ends with a mild big crunch where, again, the
spacial volume vanishes along a different direction. The
manifold thus is not isotropic. The maximal spacial volume of the
cosmological region increases with the algebraic level, $k$, of
the affine Lie algebra. The source of the infinite sequence of
bangs and crunches is that in order to avoid having ab-initio
closed time-like curves (CTC) in the manifold, one considers the
universal cover of $SL(2,\IR)$ and not just the group itself. One
may pose the set of questions referring to these compact regions.
The surprise is the appearance of
additional regions: the whisjers, denoted by W in figure 1.
These regions extend
all the way to different spacial infinities. Their emergence sheds
a new light on the set of questions.

\item The Whiskers Regions and the S Matrix Problem \cite{egkr}:

The W regions have infinite extent and moreover their metric is
time-independent. String theory has thus provided a place to
build accelerators so one can measure the S matrix. The common
wisdom claims one should be able to do it in string theory and
indeed one can. The whiskers also contain time-like submanifolds
which are singular and CTCs. Their significance needs to be
studied.

\item Observables and Entropy \cite{egkr}:

The BRST exact operators are natural candidates to be the
observables. Their number could be defined as the entropy of the
system. The analysis shows that the number of states is infinitely
depleted relative to the number of states in the $AdS_3$ case. In
the latter case the ten-dimensional string theory can be
represented by a two-dimensional dual conformal field theory. The
case at hand contains even fewer states \cite{egkr}. Examples of
observables that were calculated are the vertex operators in the
theory. One of them, U, corresponding to a $\delta$ function
normalizable operator, is given by \cite{egkr,dvv}:
 \eqn{vertexop}{V=UDf}
where $D$ is the $SU(2)$ part of the vertex, $f$ is the internal
$M$ part, and $U$ is given by:
 \eqn{vjm}{U(m,m';j,\epsilon;g)= K_{++}-{\sin(\pi(j+im'))
 \over\sin(\pi(j+im))}K_{--}}
where $m,m',j$ are the
usual $SL(2)$ numbers related to the energy and momenta in target space
(see below), and $K_{++}$, $K_{--}$ are defined in \cite{vil}.
What is important
here is that $U$ has the following asymptotic form on the boundary
of the whiskers:
 \eqn{asu}{U(E_{\pm};j;\theta_2\to\infty)\sim
e^{2iE_-\phi}e^{-\theta_2}\Big[ e^{2i(E_+ t+p_s\theta_2)}
+R(j;m,m')e^{2i(E_+ t-p_s\theta_2)} \Big]} where
\eqn{omch}{E_{\pm}=\half(m\pm m'), \quad j= -{1 \over 2}+ip_s}
{}From this form one can extract the reflection coefficient of
each partial wave. Its square is given by: \eqn{R}{1\geq
|R|^2={\cosh (2\pi E_+) +\cosh\( 2\pi (p_s-E_-)\) \over \cosh
(2\pi E_+) +\cosh\( 2\pi (p_s+E_-)\) }} The theory is unitary.
This results from the fact that string theory on $SL(2,\IR)$ is
unitary and has similar lower point functions. The interpretation
of that part of the vertex operator as a reflection coefficient
was possible after choosing the appropriate boundary conditions.
The system is defined far away from the singularities in a weak
coupling regime as one would indeed prefer.

\item Bounday Conditions and Possible Crossing of the Singularities:

Once the boundary conditions are set in a weakly coupled region,
the vertex operator is fixed everywhere. This includes the regions
following the particular big bang and the following big
crunches and big bangs. A projectile starting in a definite
partial wave in one of the whiskers ends up also in other
whiskers. Again, unitarity of string theory on $SL(2,\IR)$ is
responsible for the unitarity when summed over all the boundaries
of the whiskers. One resolution of the problem of crossing the
singularity would be that the singularity was there only
semi-classically. Had the singularities had no algebraic origin
that would be likely. However, one can associate some of the
singularities which are manifest in the geometrical formulation
with fixed points of the algebraic identifications implied in
constructing the coset. Other are not and can be removed as will
be discussed later.

\item Holographic Aspects:

At this stage one would still require that holography be derived
for each case and not be added to the axioms of a theory of
gravity. There is thus no guarantee that this background has a
holographic description, nevertheless one can observe several
primodial seeds of this feature. The universe seems to be divided
in cosmologies and whiskers. Holographic screens, if they have a
meaning, are likely to be set at the asymptotic boundaries of the
whiskers. ``We'' on the other hand are presumably situated in the
cosmological parts. From the point of view of the bulk theory
there are several types of representations of $SL(2,\IR)$ which
play a role in the mapping between the boundary and the
bulk theories. The discrete representations of $SL(2,\IR)$
reflect normalizable states which live near the cosmological
region, the information about them may be encoded in the
non-normalizable operators from that series which extend to the
boundary. The principal continuous representations give more
global information about the physics in the bulk and are related
on the boundary to the $\delta$ function normalizable operators.
The system seems to have an infinite number of screens. Thus one
may compose the dual theory out of an infinite entangled Hilbert
spaces, the wave function being a product of an infinite number of
correlated states (a general claim on this issue was made in
\cite{maldacena}). Recently, in the BTZ black hole case a
reduction in the number of effective boundaries was indicated
\cite{kos}. This could be however related to special scattering
properties of the black hole. As mentioned strings propagating on
$AdS_3$ times an attendant manifold do have duals
\cite{JMMaldacena}. One may expect that the same is true after
gauging and appropriate compensation of the central charge. Such a
non-perturbative definition could be of value, and studies of the
simpler BTZ black hole case were done. To summarize the positive
results, the study has allowed the explicit calculation of some
interesting properties of strings in the presence of cosmology and
has offered some surprising ways to resolve long standing problems
in GR. There are however also many open issues.

\item Back Reaction and Other Open Issues:

As described earlier the cosmological background is exactly of
such a complicated nature that it could be destabilized by a speck
of dust. The worry that string perturbation theory would
invalidate the supposedly exact calculation at tree level arises
because the various exact solutions have CTC and less generic
closed null-like curves, in addition to the space-like
singularities. We have reviewed the field theory calculation
pointing at a singular energy density at the singularity. The hope
was that a string probe smears and thus heals the singularity. The
analysis of the lower point functions gives indeed finite answers.
This is not the case however for the four point function in one of
the orbifold cases; it diverges \cite{lms1}. The source of that
divergence is that the Hamiltonian of the system does not commute
with the projection operator of the orbifold construction. As all
states must be singlets of that projection, the allowed states can
not be eigne-states of energy. There is no way to construct a
solely low energy scattering probe. The state needs to contain
higher energy components. The question becomes, what is the weight
of these high energy components? For the original case it was
found that in fact all components had equal weight and thus the
average energy of the projectile was infinite, leading to a
breakdown of string perturbation theory. In other cases a control
parameter was found that allowed to obtain non-singular scattering
\cite{egr,lms2,fs}. For the singularity in the coset the four
point function has not yet been calculated exactly. On top of
that, non-perturbative effects, such as black hole formation in a
region where GR calculations are reliable also threaten the
validity of perturbation theory \cite{hp}. Preliminary results for
the BTZ black hole case did not detect a singularity \cite{kos}.
The coset case has actually many similarities to the BTZ case
\cite{egr}. However, these results are not yet conclusive and the
issue remains open. The various theories can be studied in a
world-volume supersysmmetic version. Some orbifolds can be
arranged to be supersymmetric also in target space. As mentioned,
the importance of this is not yet clear. Also long strings and
various thermodynamical aspects have not yet been studied. There
has been some initial work on D-brane probes. The branes probe
shorter distances and may offer an additional view on how strings
experience the singularities. An uninvited feature emerged both in
the compact cosmological backgrounds and several of the orbifold
constructions. The systems have regions which contain CTCs. The
first negative associations with such curves is that they allow
the violation of causality. In the gauged models the origin of the
CTCs is the compact $SU(2)$ manifold. Gauge invariance does
however select only vertex operators which are single valued. The
history they represent repeating itself. In fact the depletion of
the number of states has the same origin and is welcome. One may
suggest that the two are correlated. The reduction of degrees of
freedom going hand in hand with those few states for which
causality, as well as other properties, remain intact. In GR CTCs
also lead to accumulation of very large energies. In some cases it
clearly invalidates perturbation theory taking away the value of
supposedly exact solutions. One is thus led to analyze the
singularity structure in even more detail. One considers control
parameters to study the singularities one of them is the radius,
$R$, of an extra fifth dimensions (Such extra parameters can be
obtained, for instance, by $O(d,d)$ rotations as in \cite{gp,gmv}
(for a review, see \cite{GPR})).

\end{itemize}

\section{Example: 2-d Cosmology}

In the previous sections we reviewed several examples of exact
time-dependent backgrounds in string theory. In particular, four
dimensional cosmologies based on Abelian quotients of $SL(2)\times
SU(2)$ were studied in detail in \cite{egkr} and discussed in
previous sections. In ref. \cite{egr} an Abelian gauge field was
turned on by considering a family of Abelian quotients of
$SL(2)\times SU(2)\times U(1)$. The presence of the background
gauge field changes the structure of the singularities. For
instance, a big bang/crunch curvature singularity and the
time-like domain wall attached to it can be ``pushed'' towards the
boundary of the whisker by turning on such a gauge field, leaving
behind a BTZ-like singularity (for details, see \cite{egr}). The
fact that such a gauge field background is still described within
the context of an exact CFT -- $SL(2)\times SU(2)\times
U(1)/U(1)^2$ -- allows to study some properties of the theory,
discussed in previous sections, along the lines of \cite{egkr}.
For instance, it was shown that uncharged incoming waves from a
whisker can be fully reflected if and only if a big bang/crunch
singularity exists, from which it is scattered.

In ref. \cite{ckr}, a two dimensional cosmology based on an
Abelian quotient of $SL(2)$ with a BTZ identification
was studied in detail along the lines of \cite{egkr}. In order to
understand better the singularity in that space we turn on an
Abelian background gauge field by considering instead an
Abelian quotient of $SL(2)\times U(1)/Z$. Below we shall
mainly consider the geometry of the two dimensional time-dependent
backgrounds obtained in this way. The consideration of observables
in such geometries can be easily obtained along the lines of
\cite{egkr,egr,ckr}.

Explicitly, in this example we construct a family of
$3$-dimensional time-dependent backgrounds by gauging the WZW
model of the $4$-dimensional $SL(2,\IR)_{k<0} \times U(1)$ group
manifold by a family of non-compact time-like $U(1)$ subgroups.
Then by taking a small constant radius of the $U(1)$ part we
obtain a two dimensional cosmology via the Kaluza-Klein (KK)
mechanism. All two dimensional spaces have: compact static
regions, which admit CTC after the BTZ identification,
non-compact time-dependent regions that are flat in the asymptotic
infinite past or future, a non-trivial dilaton and a background KK
gauge field. All regions are generically separated by horizons, which turn
into orbifold singularities after the $Z$ identification.
This family of backgrounds admits two kinds of singularities: one
is generated by fixed points of the gauge group (and lie in the
compact region), and the other by fixed points of an orbifold
identification (and lie between regions). These backgrounds can
also be obtained by $O(1,2)\subset O(2,2)$ rotations along the
lines of \cite{GR} (for a review, see \cite{GPR}). In that family,
a specific two dimensional space, with a vanishing gauge field,
was studied in \cite{ckr}. In this section we shall concentrate on
the geometry of those two dimensional cosmologies with a
background gauge field.

Let $(g,x)\in SL\(2\)\times U\(1\)$ be a point on the product
group manifold where $x\sim x+2\pi r$ and let $-\kappa\equiv k<0$
be the level of $SL(2,\IR)$. The $U(1)$ gauge group acts as
 \eqn{gatr}{(g,x_L,x_R)\rightarrow (e^{\rho
\sigma_{3}/\sqrt{\kappa} }g e^{\tau \sigma_{3}/\sqrt{\kappa}},
 x_L+\rho',x_R+\tau')~.}
Since we gauge only $U(1)$ out of the two right-handed $U(1)$
generators in \gatr, the two parameters
$(\tau,\tau')\equiv\underline\tau$ are not independent but rather
are constrained by
 \eqn{constr}{\underline\tau\equiv\tau\underline u~,}
where $\underline u$ is some unit real 2-vector. The left-handed
parameters $(\rho,\rho')\equiv\underline\rho$ in \gatr\ depend
linearly on the right-handed $\underline\tau$ parameters. For an
anomaly free gauging this dependence has to take the form
 \eqn{orth}{\underline{\rho} = R \underline\tau~, }
where the matrix $R$ is an $O(1,1;\IR)$ matrix~\footnote{For
$SL(2,\IR)$ with a positive level simply substitute $i\alpha$ for
$\alpha$ in every equation.}
\eqn{rmatrix}{R=\left(\matrix{\ch(\alpha)&\sh(\alpha)\cr
 \sh(\alpha)&\ch(\alpha)}\right).}
The gauged action, as in \cite{egr}, is then defined by
 \eqn{gact}{S=S[e^{\hat\rho \sigma_{3}/\sqrt{\kappa} }g
e^{\hat\tau \sigma_{3}/\sqrt{\kappa}}]+
S'[x+\hat \rho'+\hat \tau']- {1\over {2\pi}}\int d^2z
(\partial\hat{ \underline {\rho}} -R\partial\hat {\underline{
\tau}})^T (\bar{\partial} \hat{\underline {\rho}} -R\bar
 {\partial} \hat{\underline {\tau}})~.}
Here, $S[g]$ is the WZW action,
 \eqn{wzg}{S[g]=-{\kappa \over {4
\pi}}[\int_{\Sigma} Tr(g^{-1}\partial g g^{-1}\bar \partial g)-{1
 \over 3} \int_{B} Tr (g^{-1}dg)^3 ]~,}
where $\Sigma$ is the string's worldsheet and $B$ a 3-submanifold
of the group $SL(2)$ bounded by the image of $\Sigma$. $S'[x]$ is
 \eqn{uone}{S'[x]={1 \over {2 \pi}}\int_{\Sigma}\d x \bar \d x~.}
Apart from the constraints \orth, $\hat{\underline\rho}$ and
$\hat{\underline\tau}$ are independent fields. The action \gact\
is invariant under the gauge transformation \gatr\ for the fields
$g$ and $x$ together with the field transformation
 \eqn{agatr}{\eqalign{&\hat{ \underline{\rho}}\rightarrow \hat
{\underline{\rho}}-\underline {\rho} \cr&\hat{\underline{
\tau}}\rightarrow \hat {\underline{\tau}}-\underline{\tau}}}
provided that the parameters $\underline{\rho}$ and
$\underline{\tau}$ satisfy the relation \orth. Using the
Polyakov-Wiegmann identity one sees that the action \gact\ depends
on $\hat{\underline{\rho}}$ and $\hat{\underline{\tau}}$ only
through the quantities
 \eqn{af}{\eqalign{&A ={\underline
u}^T\partial\hat{\underline{\tau}}\cr &\bar{A}=(R\underline
 u)^T\cdot\bar{\partial}\hat{\underline{\rho}}}}
The gauged action has then the form
 \eqn{act}{S= S[g]+S'[x]+{1 \over {2\pi}}\int
d^2z[ A\bar{\bf J}^T \cdot\underline u +\bar{A}{\bf J}^T \cdot
R\underline u +2A\bar{A}{\underline u}^T M \cdot R \underline u]~.}
$A$ and $\bar A$ are holomorphic and anti-holomorphic gauge
fields. ${\bf J}^T$ and $\bar {\bf J}^T$ are the row vector of
currents,
 \eqn{cur}{\eqalign{{\bf J}^T&= (\sqrt{\kappa}Tr[ \partial g
g^{-1}\sigma_3], 2\d x) \cr \bar{\bf J}^T&= (\sqrt{\kappa}
 Tr[g^{-1} \bar{\partial} g\sigma_3], 2\bar{ \partial} x)}}
The $2\times 2$ matrix $M$ in \act\ is of the form,
 \eqn{qfo}{M=\left(\matrix{{1\over 2}Tr[g^{-1}\sigma_3 g\sigma_3]
 &0\cr 0&1}\right) + R~. }
The scalar multiplication $(\cdot)$ is the one in $O(1,1)$, so for
example ${\underline v}^T \cdot\underline v \equiv -v_1^2+v_2^2$.
One can write the same action as a complete square
 \eqn{actsq}{\eqalign{S=& S[g]+S'[x]+
\cr +&{1 \over {2\pi}}\int d^2z\[\(a+{\bar{\bf J}^T
\cdot\underline u \over 2{\underline u}^T M \cdot R \underline
u}\)2{\underline u}^T M \cdot R \underline u \(\bar a+{{\bf J}^T
\cdot R\underline u\over 2{\underline u}^T M \cdot R \underline
u}\)-{\(\bar{\bf J}^T \cdot\underline u \)\({\bf J}^T \cdot
R\underline u\) \over 2{\underline u}^T M \cdot R \underline
 u}\]}}
After integrating out the fields $A$ and $\bar A$, one gets, to
the first order in ${1\over \kappa}$, the action
 \eqn{actaia}{S=S[g]+S'[x]-{1 \over {4\pi}}\int d^2z\[{\(\bar{\bf J}^T
\cdot\underline u \)\({\bf J}^T \cdot R\underline u\) \over
 {\underline u}^T M \cdot R\underline u}\]}
and the dilaton becomes
 \eqn{dil}{\Phi = \Phi_0-{1\over 2}\log\({\underline u}^T M
 \cdot R\underline u\)~.}
The gauge invariance of the action is fixed by setting
 \eqn{gaugefix}{g=e^{{1\over 2}y\sigma_3}g(\theta_i)e^{-{1\over 2}y\sigma_3}~.}
The definition of the factor $g(\theta_i)$ depends on the region
where $g$ is in the $SL(2)$ group manifold \cite{egkr,vil}.
Defining
 \eqn{w}{W=Tr(\sigma_3 g \sigma_3 g^{-1})~,}
 $g(\theta_1)$ stands for $e^{i\theta_1 \sigma_2}$ in regions of $SL(2)$
for which $W$ satisfies $|W|\le 2$. The points of $SL(2)$ for
which $W>2$ are divided into $4$ regions. There the factor
$g(\theta_2)$ represents
 $\pm e^{\pm\theta_2 \sigma_1}$. For the $4$ regions where
$W<-2$, $g(\theta_3)=\pm i\sigma_2 e^{\pm\theta_3 \sigma_1}$. At
the point $\theta_1=0$, $W=2$. Here two of the regions
parameterized by $\theta_2$ meet the region parameterized by
$\theta_1$. Similarly, at $\theta_1=\pi$ the other two regions
parameterized by
 $\theta_2$ meet the region parameterized by $\theta_1$.
At $\theta_1={\pi \over 2}$ $(W=-2)$ two regions parameterized by
$\theta_3$ meet the $\theta_1$ region and at $\theta_1={3\pi \over
2}$ the other two $\theta_3$ regions meet the $\theta_1$ region
(see figure 3).
The range of $\theta_{2,3}$ is $0\le \theta_{2,3} < \infty$. For
the group $SL(2)$, $\theta_1$ satisfies $0 \le \theta_1 \le 2\pi$.
For the infinite cover of $SL(2)$, $\theta_1$ satisfies $-\infty <
\theta_1 < \infty$. For $PSL(2)$ (or the \Poincare patch), $0 <
\theta_1 \leq \pi$.

After plugging \gaugefix\ into \actaia, \dil\ one gets
 \eqn{finalact}{\eqalign{S=&{1 \over 2\pi}\int_{\Sigma}\d x \bar \d
x+{\kappa\over 2\pi}\int
d^2z\[\d\theta_1\bar\d\theta_1-\sin^2(\theta_1)\d y\bar\d y\]+ \cr
+&{1\over \pi}\int d^2z{\(\sqrt{\kappa}\sin^2(\theta_1){\underline
u}_1\bar\d y+{\underline u}_2\bar\d x\)
\(\sqrt{\kappa}\sin^2(\theta_1)(R\underline u)_1\d y-(R\underline
u )_2\d x\) \over {\underline u}^T M \cdot R\underline u} }}
\eqn{finaldil}{\Phi = \Phi_0-{1\over 2}\log\({\underline u}^T M
 \cdot R \underline u \)}
where $|W|\le 2$. In the regions where $W>2$, $\theta_1$ in
\finalact, \finaldil\ should be replaced by $i\theta_{2}$. In the
regions with $W<-2$, substitute $i\theta_3$ for $\theta_1 -
{\pi\over 2}$.

If we take the vector ${\underline u}^T= (1,0)$ then $G_{x,x}$ is
constant~\footnote{Actually, $G_{x,x}=const$ iff $(G+B)_{y,x}=0$ and,
therefore, in this case the ${SL(2)\times U(1)\over U(1)}$ background can
be used in the heterotic string.}
and after resealing $x \rightarrow \sqrt{\kappa}x$ the
action and the dilaton becomes
 \eqn{finalactu}{\eqalign{S=&{1 \over
2\pi}\int_{\Sigma}\d x \bar \d x+{\kappa\over 2\pi}\int d^2z
\[ \d \theta_1\bar \d \theta_1-\sin^2(\theta_1)\d y\bar\d y \]- \cr
-&{1\over \pi}\int d^2z{\sqrt{\kappa}\sin^2(\theta_1)\bar\d y
\(\sqrt{\kappa}\sin^2(\theta_1)\ch(\alpha)\d y-\sh(\alpha)\d x\)
\over 1+\ch(\alpha)\cos(2\theta_1)}= \cr =&{\kappa\over 2\pi}\int
d^2z\[\d\theta_1\bar\d\theta_1- {\d y\bar\d y -2\th({\alpha\over
2})\bar\d y \d x \over \cot^2(\theta_1)-\th^2({\alpha\over 2})} +
\d x \bar \d x\] }}
 \eqn{finaldilu}{\Phi = \tilde{\Phi}_0-{1\over
2} \log\(\cos^2(\theta_1)-\th^2({\alpha\over
 2})\sin^2(\theta_1)\)}
Again, in regions for which $|W|>2$ make the appropriate
replacement for $\theta_1$. For large $\kappa$ and for small
radius of the circle parameterized by $x$, this action \finalactu\
describes a $2$-dimensional time-dependent space-time
parameterized by $(\theta_i,y)$. The $3$-dimensional metric and
antisymmetric tensor read from \finalact\, produce a corresponding
$2$-dimensional structure via the Kaluza-Klein mechanism. The term
proportional to $\d x \bar{\d} y$ gives rise in two dimensions to
a $U(1)$ gauge field whose charge is the momentum as well as the
winding along the $x$ circle. The $2$-dimensional metric and
background gauge field take the form
 \eqn{twodimbg}{\eqalign{{1\over
\kappa}ds^2&=d\theta_1^2-{\cot^2(\theta_1) \over
\(\cot^2(\theta_1) -\th^2({\alpha\over 2})\)^2}\d y\bar\d y \cr
A_y&={\sqrt{\kappa}\th({\alpha\over 2})\over
 \cot^2(\theta_1)-\th^2({\alpha\over 2})} }}
In regions for which $|W|>2$ make the appropriate replacement for
$\theta_1$.~\footnote{The charged 2-d black hole
${SL\(2\)_{k>0}\times U\(1\) \over U\(1\)_{space-like}}$ is the
same as \finalactu, \finaldilu, \twodimbg\ with $\alpha\rightarrow
i\alpha$, $x\rightarrow ix$ and $\kappa\rightarrow -\kappa$.}

\begin{figure}[h] \begin{center}
\epsfxsize = 6.5in \epsffile{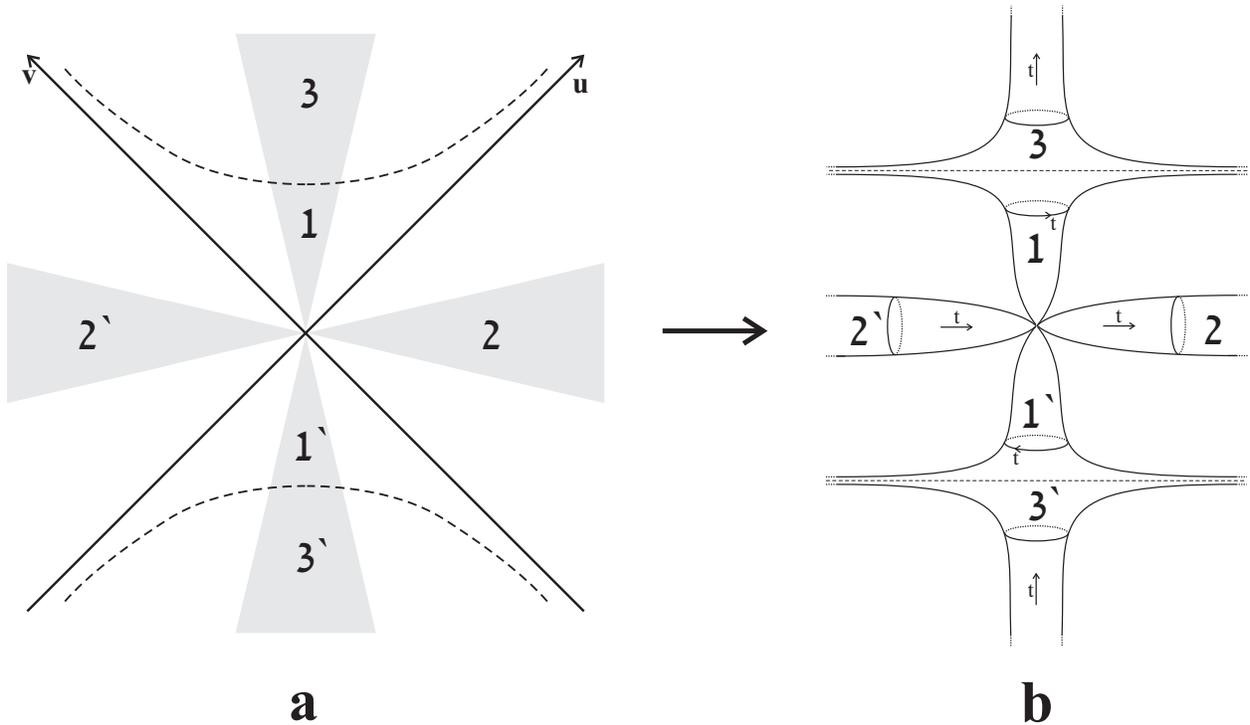} \caption{{\bf a}: The
two dimensional cosmology where the fundamental domains are
marked; the dashed lines are curvature singularities. {\bf b}: The
space after the identification.} \label{fig 3} \end{center}
\end{figure}

In the Kruskal coordinates~\footnote{The coordinates $u$
and $v$ cover all ${PSL(2)\over U(1)}$, and only half of
${SL(2)\over U(1)}$ (or the \Poincare patch of the universal
cover).} (for $|W|\le 2$)
 \eqn{uvcord}{u=\sin(\theta_1)e^y~, \quad v=\sin(\theta_1)e^{-y}}
the metric, dilaton and gauge field (\twodimbg ,\finaldilu) are:
 \eqn{twodimbguv}{\eqalign{{1\over \kappa}ds^2&={v^2du^2 +u^2dv^2\over
4uv}\( {1\over 1-uv}-{1-uv\over \(1-uv[1+\tanh^2({\alpha\over
2})]\)^2}\) \cr &+{dudv\over 2}\( {1\over 1-uv}+{1-uv\over
\(1-uv[1+\tanh^2({\alpha\over 2} )]\)^2}\)}}
 \eqn{uvdilaton}{\Phi=\Phi_0-\half \log \(1-uv[1+\tanh^2({\alpha\over 2})]\)}
 \eqn{uvA}{A_u={\sqrt{\kappa}\over2}{v\th({\alpha\over
 2})\over 1-uv\(1+\tanh({\alpha\over 2})\)},\quad A_v=-{\sqrt{\kappa}\over2}
{u\th({\alpha\over
 2})\over 1-uv\(1+\tanh({\alpha\over 2})\)}}
For the degenerate case $\alpha=0$, the metric, dilaton and
background gauge field can be written as follows
 \eqn{twodbh}{\eqalign{ds^2&=-\kappa{dudv\over
 1-uv} \cr \phi&=-{1\over 4}\log (1-uv)^2} \quad A_u=A_v=0}
This is the two dimensional Lorentzian black hole background with a negative
level.

So far we have constructed ${SL(2) \times U(1) \over \IR}$. To get
${SL(2) \times U(1) \over U(1)\times Z}$ we further identify
 \eqn{orbifuld}{g\sim
e^{\pi\lambda\sigma_3}ge^{-\pi\lambda\sigma_3} \quad
 \Longleftrightarrow \quad y\sim y+2\pi\lambda}
The identification in $SL(2)$ is the one which leads to the BTZ
black hole background (with a negative level), so our space is
${BTZ_{k<0} \times U(1) \over U(1)}$.

In the degenerate case $\alpha=0$, \orbifuld\ becomes $(u,v)\sim
(ue^{-\lambda},ve^\lambda)$. This space was recently studied in
\cite{ckr} and is plotted in figure 3. Regions 2,2',3,3' are
time-dependent, and approach flat space at early or late times.
Regions 1,1' are static and have closed time-like curves. Regions
1,1',2,2' meet at an orbifold singularity which is locally an
$R^{1,1}$ modded by boost singularity. Regions 1 and 3 meet at a
curvature singularity which coincides with an orbifold
singularity, and the same for regions 1' and 3'. The lines $u=0$ and $v=0$
in figure 3{\bf a} lead to a non-Hausdorf structure
which is not indicated in figure 3{\bf b}.

\begin{figure}[h] \begin{center}
\epsfxsize = 6.5in \epsffile{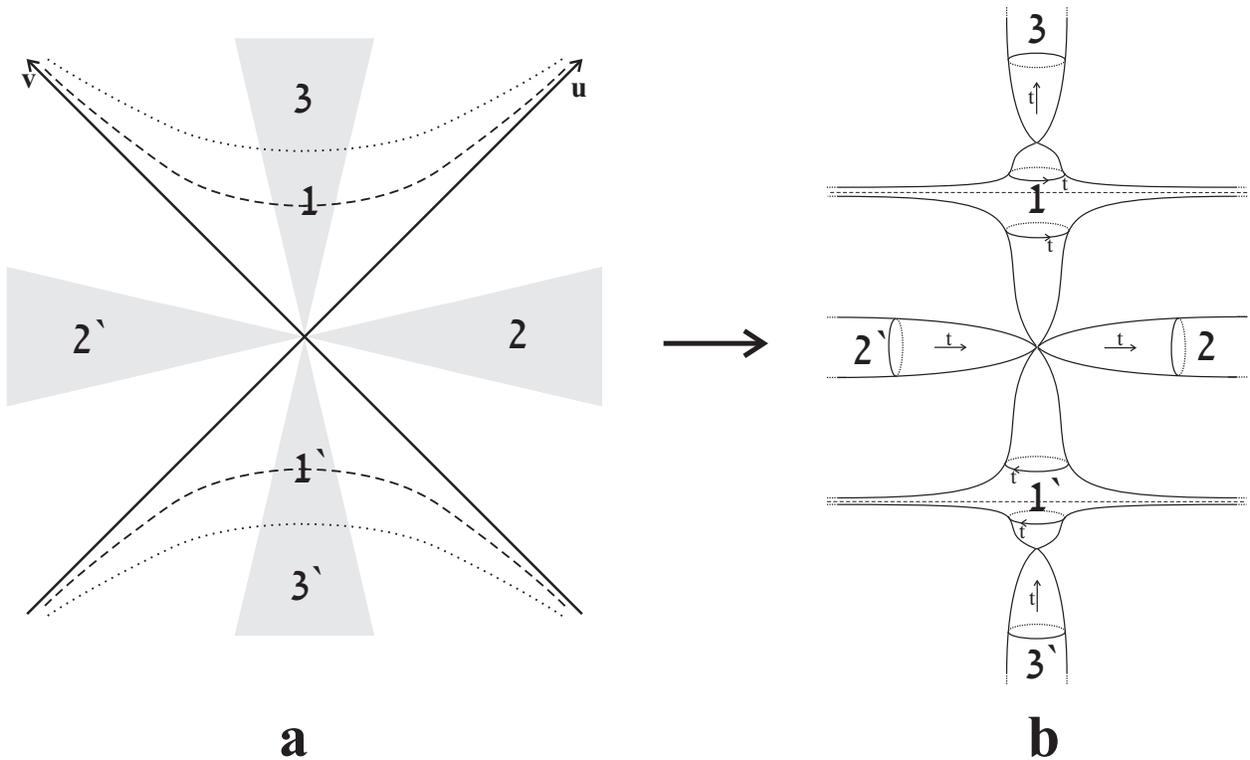} \caption{{\bf a}: The
2-dimensional cosmology with a background gauge field ($\alpha\neq
0$); the dashed lines are curvature singularities and the doted
lines are horizons. {\bf b}: The space after the identification;
the dashed lines are time-like curvature singularities -- domain
walls -- in the static compact regions. The horizons in {\bf a}
turned into big bang/crunch orbifold singularities, connecting the
various regions.} \label{fig 4} \end{center} \end{figure}

When we turn on $\alpha\neq 0$, the curvature singularities move
into the compact static regions (1 and 1').~\footnote{In the case
of a positive level $k>0$ (when the resulting quotient is a
charged black hole), the curvature singularity is moved into what
is now a static, non-compact region.} In ${SL(2) \times U(1) \over
U(1)\times Z}$ it leaves an orbifold singularity behind (see
figure 4). It is an orbifold singularity since at
$\theta_1={\pi\over 2}$ one can do a gauge transformation together
with the orbifold identification \orbifuld\, such that it acts
only in the $U(1)$ part and, therefore, it becomes a fixed point
after KK reduction. If we where considering the space
with no orbifold identification or KK reduction, the surface
$\theta_1={\pi\over 2}$ would have been a horizon instead of
being a singularity.

To summarize, we considered a two dimensional cosmology with a
positive cosmological constant, in the presence of an Abelian
gauge field. As in \cite{egr}, we find that turning on an Abelian
gauge field in the background studied in \cite{ckr}, changes the
structure of the singularities. The singularities split
into a curvature singularity and orbifold singularities. The
curvature singularities become time-like, and are located inside
the compact static regions. On the other hand, the orbifold
$R^{1,1}/$Boost--like singularities are big bang and crunch
singularities of the expending and contracting universes,
respectively. Unlike ref. \cite{egr}, where the curvature
singularities can be removed completely by tuning the gauge field
such that they are pushed towards the boundaries of the whiskers
in figure 1, here the singularities cannot be removed. They are
``stuck'' inside the compact static regions.

The fact that this background is an ${SL(2)\times U(1)\over U(1)\times Z}$
CFT sigma-model allows to extract some exact results. For
instance, the Bogolubov coefficients computed in \cite{ckr} eq.
(4.32) apply to our $\alpha$-family as well; only the dispersion
relation is modified as a function of the $\alpha$-dependent gauge
condition. As a consequence, particle creation remains for a
generic gauge field (eq. (4.37) in \cite{ckr} is valid here as
well).

\bigskip\noindent {\bf Acknowledgements:} We are grateful to S.~Elitzur and
D.~Kutasov for collaborations. This work is supported in part by BSF --
American-Israel Bi-National Science Foundation, the Israel Academy
of Sciences and Humanities -- Centers of Excellence Program, the
German-Israel Bi-National Science Foundation, the European RTN
network HPRN-CT-2000-00122, and the Horwitz foundation (AS).



\begin{thebibliography}{77}

\bibitem{Israel}
For example: W.~Israel, ``Thermo Field Dynamics Of Black Holes,''
Phys.\ Lett.\ A {\bf 57}, 107 (1976).

\bibitem{Bousso}
W.~Fischler and L.~Susskind, ``Holography and cosmology,''
arXiv:hep-th/9806039. R.~Bousso, ``A Covariant Entropy
Conjecture,'' JHEP {\bf 9907}, 004 (1999) [arXiv:hep-th/9905177].

\bibitem{Banks}
J.~M.~Maldacena and A.~Strominger, ``Statistical entropy of de
Sitter space,'' JHEP {\bf 9802}, 014 (1998) [arXiv:gr-qc/9801096].
S.~Hawking, J.~M.~Maldacena and A.~Strominger, ``DeSitter entropy,
quantum entanglement and AdS/CFT,'' JHEP {\bf 0105}, 001 (2001)
[arXiv:hep-th/0002145]. T.~Banks, ``Cosmological Breaking Of
Supersymmetry?,'' Int.\ J.\ Mod.\ Phys.\ A {\bf 16}, 910 (2001).

\bibitem{hh}
see for example: S.~W.~Hawking, ``The Chronology protection
conjecture,'' Phys.\ Rev.\ D {\bf 46}, 603 (1992).

\bibitem{hk}
W.~A.~Hiscock and D.~A.~Konkowski, ``Quantum Vacuum Energy In Taub
- Nut (Newman-Unti-Tamburino) Type Cosmologies,'' Phys.\ Rev.\ D
{\bf 26}, 1225 (1982).

\bibitem{egkr}S.~Elitzur, A.~Giveon, D.~Kutasov and E.~Rabinovici,
``From big bang to big crunch and beyond,''
JHEP {\bf 0206}, 017 (2002)
[arXiv:hep-th/0204189].

\bibitem{egr}
S.~Elitzur, A.~Giveon and E.~Rabinovici, ``Removing
singularities,'' JHEP {\bf 0301}, 017 (2003)
[arXiv:hep-th/0212242].

\bibitem{ess}
P.~Horava, ``Some exact solutions of string theory in
four-dimensions and five-dimensions,'' Phys.\ Lett.\ B {\bf 278},
101 (1992) [arXiv:hep-th/9110067]. C.~Kounnas and D.~Lust,
``Cosmological string backgrounds from gauged WZW models,'' Phys.\
Lett.\ B {\bf 289}, 56 (1992) [arXiv:hep-th/9205046].
A.~A.~Tseytlin, ``Exact string solutions and duality,''
arXiv:hep-th/9407099. J.~Simon, ``The geometry of null rotation
identifications,'' arXiv:hep-th/0203201. A.~J.~Tolley and
N.~Turok, ``Quantum fields in a big crunch / big bang spacetime,''
arXiv:hep-th/0204091.

\bibitem{nek}
N.~A.~Nekrasov, ``Milne universe, tachyons, and quantum group,''
arXiv:hep-th/0203112.

\bibitem{cc}
L.~Cornalba and M.~S.~Costa, ``A New Cosmological Scenario in
String Theory,'' Phys.\ Rev.\ D {\bf 66}, 066001 (2002)
[arXiv:hep-th/0203031].

\bibitem{gp}
A.~Giveon and A.~Pasquinucci, ``On cosmological string backgrounds
with toroidal isometries,'' Phys.\ Lett.\ B {\bf 294}, 162 (1992)
[arXiv:hep-th/9208076].

\bibitem{bs}
I.~Bars and K.~Sfetsos, ``$SL(2,\IR) \times SU(2) / \IR^2$ string
model in curved space-time and exact conformal results,'' Phys.\
Lett.\ B {\bf 301}, 183 (1993) [arXiv:hep-th/9208001].

\bibitem{nw}
C.~R.~Nappi and E.~Witten, ``A Closed, expanding universe in
string theory,'' Phys.\ Lett.\ B {\bf 293}, 309 (1992)
[arXiv:hep-th/9206078].

\bibitem{lms1}
H.~Liu, G.~Moore and N.~Seiberg, ``Strings in a time-dependent
orbifold,'' JHEP {\bf 0206}, 045 (2002) [arXiv:hep-th/0204168].

\bibitem{lms2}
H.~Liu, G.~Moore and N.~Seiberg, ``Strings in time-dependent
orbifolds,'' JHEP {\bf 0210}, 031 (2002) [arXiv:hep-th/0206182].

\bibitem{ckr}B.~Craps, D.~Kutasov and G.~Rajesh, ``String
propagation in the presence of cosmological singularities,'' JHEP
{\bf 0206}, 053 (2002) [arXiv:hep-th/0205101].

\bibitem{afhs}
for example see: O.~Aharony, M.~Fabinger, G.~T.~Horowitz and
E.~Silverstein, ``Clean time-dependent string backgrounds from
bubble baths,'' JHEP {\bf 0207}, 007 (2002)
[arXiv:hep-th/0204158].

\bibitem{sen}
see for example: A.~Sen, ``Time evolution in open string theory,''
JHEP {\bf 0210}, 003 (2002) [arXiv:hep-th/0207105].

\bibitem{law}
A.~Lawrence, ``On the instability of 3D null singularities,'' JHEP
{\bf 0211}, 019 (2002) [arXiv:hep-th/0205288].

\bibitem{maldacena}
J.~M.~Maldacena, ``Eternal black holes in Anti-de-Sitter,''
arXiv:hep-th/0106112.

\bibitem{whp}
S.~W.~Hawking and D.~N.~Page, ``Thermodynamics Of Black Holes In
Anti-De Sitter Space,'' Commun.\ Math.\ Phys.\  {\bf 87}, 577
(1983). E.~Witten, ``Anti-de Sitter space, thermal phase
transition, and confinement in  gauge theories,'' Adv.\ Theor.\
Math.\ Phys.\ {\bf 2}, 505 (1998) [arXiv:hep-th/9803131].

\bibitem{orbi}
A.~Giveon and A.~D.~Shapere, ``Gauge symmetries of the N=2
string,'' Nucl.\ Phys.\ B {\bf 386}, 43 (1992)
[arXiv:hep-th/9203008]. G.~W.~Moore, ``Finite In All Directions,''
arXiv:hep-th/9305139. J.~Khoury, B.~A.~Ovrut, P.~J.~Steinhardt and
N.~Turok, ``Density perturbations in the ekpyrotic scenario,''
Phys.\ Rev.\ D {\bf 66}, 046005 (2002) [arXiv:hep-th/0109050].
V.~Balasubramanian, S.~F.~Hassan, E.~Keski-Vakkuri and A.~Naqvi,
``A space-time orbifold: A toy model for a cosmological
singularity,'' Phys.\ Rev.\ D {\bf 67}, 026003 (2003)
[arXiv:hep-th/0202187]. E.~J.~Martinec and W.~McElgin, ``Exciting
AdS orbifolds,'' arXiv:hep-th/0206175. M.~Fabinger and
J.~McGreevy, ``On smooth time-dependent orbifolds and null
singularities,'' arXiv:hep-th/0206196. J.~Simon, ``Null orbifolds
in AdS, time dependence and holography,'' JHEP {\bf 0210}, 036
(2002) [arXiv:hep-th/0208165].

\bibitem{fs}
J.~Figueroa-O'Farrill and J.~Simon, ``Supersymmetric Kaluza-Klein
reductions of M2 and M5 branes,'' arXiv:hep-th/0208107.

\bibitem{bh_go}
K.~Bardakci and M.~B.~Halpern, ``New Dual Quark Models,'' Phys.\
Rev.\ D {\bf 3}, 2493 (1971). P.~Goddard, A.~Kent and D.~I.~Olive,
``Virasoro Algebras And Coset Space Models,'' Phys.\ Lett.\ B {\bf
152}, 88 (1985).

\bibitem{zamo}
A.~B.~Zamolodchikov, ``'Irreversibility' Of The Flux Of The
Renormalization Group In A 2-D Field Theory,'' JETP Lett.\  {\bf
43}, 730 (1986) [Pisma Zh.\ Eksp.\ Teor.\ Fiz.\ {\bf 43}, 565
(1986)].

\bibitem{subgauge}
K.~Bardakci, E.~Rabinovici and B.~Saering, ``String Models With
$C<1$ Components,'' Nucl.\ Phys.\ B {\bf 299}, 151 (1988).
D.~Altschuler, K.~Bardakci and E.~Rabinovici, ``A Construction Of
The $C<1$ Modular Invariant Partition Functions,'' Commun.\ Math.\
Phys.\  {\bf 118}, 241 (1988). W.~Nahm, ``Gauging Symmetries Of
Two-Dimensional Conformally Invariant Models,'' UCD-88-02.
D.~Karabali, Q.~H.~Park, H.~J.~Schnitzer and Z.~Yang, ``A Gko
Construction Based On A Path Integral Formulation Of Gauged
Wess-Zumino-Witten Actions,'' Phys.\ Lett.\ B {\bf 216}, 307
(1989). K.~Gawedzki and A.~Kupiainen, ``Coset Construction From
Functional Integrals,'' Nucl.\ Phys.\ B {\bf 320}, 625 (1989).

\bibitem{2dbh}
K.~Bardacki, M.~J.~Crescimanno and E.~Rabinovici, ``Parafermions
From Coset Models,'' Nucl.\ Phys.\ B {\bf 344}, 344 (1990).
E.~Witten, ``On string theory and black holes,'' Phys.\ Rev.\ D
{\bf 44}, 314 (1991).

\bibitem{1overk}
R.~Dijkgraaf, H.~Verlinde and E.~Verlinde, ``String propagation in
a black hole geometry,'' Nucl.\ Phys.\ B {\bf 371}, 269 (1992).
A.~A.~Tseytlin, ``Conformal sigma models corresponding to gauged
Wess-Zumino-Witten theories,'' Nucl.\ Phys.\ B {\bf 411}, 509
(1994) [arXiv:hep-th/9302083]. H.~J.~de Vega, A.~L.~Larsen and
N.~Sanchez, ``Non-singular string-cosmologies from exact conformal
field theories,'' Nucl.\ Phys.\ Proc.\ Suppl.\  {\bf 102}, 201
(2001) [arXiv:hep-th/0110262].

\bibitem{tdcr}
M.~Gasperini and G.~Veneziano, ``Pre - big bang in string
cosmology,'' Astropart.\ Phys.\  {\bf 1}, 317 (1993)
[arXiv:hep-th/9211021]. R.~Brustein and G.~Veneziano, ``The
Graceful exit problem in string cosmology,'' Phys.\ Lett.\ B {\bf
329}, 429 (1994) [arXiv:hep-th/9403060]. G.~Veneziano, ``String
cosmology: The pre-big bang scenario,'' arXiv:hep-th/0002094.
J.~Khoury, B.~A.~Ovrut, N.~Seiberg, P.~J.~Steinhardt and N.~Turok,
``From big crunch to big bang,'' arXiv:hep-th/0108187.

\bibitem{dvv}
R.~Dijkgraaf, H.~Verlinde and E.~Verlinde, ``String propagation in
a black hole geometry,'' Nucl.\ Phys.\ B {\bf 371}, 269 (1992).

\bibitem{vil}  N.J.~Vilenkin, ``Special Functions and the Theory
of Group Representations'' AMS, 1968; N.J.~Vilenkin, A.U.~Klimyk
``Representation of Lie Groups and Special Function'' Kluwer
Academic Publishers, 1991.

\bibitem{kos}
P.~Kraus, H.~Ooguri and S.~Shenker, ``Inside the horizon with
AdS/CFT,'' arXiv:hep-th/0212277.

\bibitem{JMMaldacena}
J.~M.~Maldacena, ``The large N limit of superconformal field
theories and supergravity,'' Adv.\ Theor.\ Math.\ Phys.\  {\bf 2},
231 (1998) [Int.\ J.\ Theor.\ Phys.\  {\bf 38}, 1113 (1999)]
[arXiv:hep-th/9711200].

\bibitem{hp}
G.~T.~Horowitz and J.~Polchinski, ``Instability of spacelike and
null orbifold singularities,'' Phys.\ Rev.\ D {\bf 66}, 103512
(2002) [arXiv:hep-th/0206228].

\bibitem{gmv}
M.~Gasperini, J.~Maharana and G.~Veneziano,
``Boosting away singularities from conformal string backgrounds,''
Phys.\ Lett.\ B {\bf 296}, 51 (1992) [arXiv:hep-th/9209052].

\bibitem{GPR} A.~Giveon, M.~Porrati and E.~Rabinovici, ``Target
space duality in string theory,'' Phys.\ Rept.\ {\bf 244}, 77
(1994) [arXiv:hep-th/9401139].

\bibitem{GR} A.~Giveon and M.~Rocek, ``Generalized duality in
curved string backgrounds,'' Nucl.\ Phys.\ B {\bf 380} (1992) 128
[arXiv:hep-th/9112070].

\end{thebibliography}
\end{document}